\documentclass[a4paper]{jpconf}
\usepackage[english]{babel}
\usepackage{graphicx,array,csquotes,biblatex}
\usepackage[pdfusetitle]{hyperref}

\usepackage{tikz}
\usetikzlibrary{chains,arrows.meta,shapes.multipart,pgfplots.groupplots}

\usepackage{listing,lstautogobble}
\definecolor{lstkeyword}{HTML}{01497c}
\definecolor{lststring}{HTML}{a4133c}
\definecolor{lstcomment}{HTML}{718355}
\lstdefinestyle{c++}{
    language=c++,
    escapechar=`,
    tabsize=3,
    backgroundcolor=\color{black!7},
    basicstyle={\ttfamily\tiny},
    keywordstyle={\color{lstkeyword}\bfseries},
    commentstyle={\color{lstcomment}},
    stringstyle={\color{lststring}},
}

\usepackage{pgfplots}
\definecolor{pgfplots_eb}{HTML}{2d00f7}
\pgfplotsset{
    compat=1.17,
    error bars/.cd,
        y dir=both,
        y explicit,
        error mark=|,
        error bar style={draw=pgfplots_eb},
        error mark options={draw=pgfplots_eb}
}
\definecolor{palette0_0}{HTML}{1f7a8c}
\definecolor{palette0_1}{HTML}{022b3a}
\pgfplotscreateplotcyclelist{twobars}{
    {draw=palette0_0,fill=palette0_0},
    {draw=palette0_1,fill=palette0_1},
}
\definecolor{palette1_0}{HTML}{fca311}
\definecolor{palette1_1}{HTML}{d00000}
\definecolor{palette1_2}{HTML}{001233}
\definecolor{palette1_3}{HTML}{00b4d8}
\definecolor{palette1_4}{HTML}{0077b6}
\pgfplotscreateplotcyclelist{perfplot}{
    {draw=palette1_0,fill=palette1_0},
    {draw=palette1_1,fill=palette1_1},
    {draw=palette1_2,fill=palette1_2},
    {draw=palette1_3,fill=palette1_3},
    {draw=palette1_4,fill=palette1_4},
}

\begin{document}
\title{RNTuple performance: Status and Outlook}
\author{Javier Lopez-Gomez$^1$, Jakob Blomer$^1$}
\address{$^1$ EP-SFT, CERN, Geneva, Switzerland}
\ead{\{javier.lopez.gomez,jblomer\}@cern.ch}

\begin{abstract}
Upcoming HEP experiments, e.g. at the HL-LHC, are expected to increase the volume of generated data by at least one order of magnitude. In order to retain the ability to analyze the influx of data, full exploitation of modern storage hardware and systems, such as low-latency high-bandwidth NVMe devices and distributed object stores, becomes critical.\par
To this end, the ROOT RNTuple I/O subsystem has been designed to address performance bottlenecks and shortcomings of ROOT’s current state of the art TTree I/O subsystem. RNTuple provides a backwards-incompatible redesign of the TTree binary format and access API that evolves the ROOT event data I/O for the challenges of the upcoming decades. It focuses on a compact data format, on performance engineering for modern storage hardware, for instance through making parallel and asynchronous I/O calls by default, and on robust interfaces that are easy to use correctly.\par
In this contribution, we evaluate the RNTuple performance for typical HEP analysis tasks. We compare the throughput delivered by RNTuple to popular I/O libraries outside HEP, such as HDF5 and Apache Parquet. We demonstrate the advantages of RNTuple for HEP analysis workflows and provide an outlook on the road to its use in production.
\end{abstract}

\section{Introduction}\label{sec:introduction}
HEP storage systems are generally tuned for write-once-read-many columnar access. Since its inception, the ROOT project\cite{brun1997root} supports the columnar storage of arbitrary C++ types and collections through TTree. However, the expected increase in the amount of experiments data that needs to be processed and the fact that TTree was not designed to make optimized use of modern hardware and storage systems, called for a new, modernized re-engineering of TTree.

RNTuple is the new, experimental, backward-incompatible ROOT columnar I/O subsystem targeting high performance, reliability, and easy-to-use robust interfaces. Despite RNTuple still being under development, at this point it is feature-complete enough to carry out an evaluation.

In this paper we contribute with the following:

\begin{itemize}
    \item A performance evaluation of TTree, RNTuple, and other well-known storage alternatives outside HEP: Apache Parquet and HDF5. Compared to a previous publication\cite{Blomer:2018icl}, we evaluate RNTuple focusing on different storage devices and a dataset with nested collections.
    \item A feature comparison and perspectives of using RNTuple in production.
\end{itemize}

\section{ROOT's RNTuple overview}\label{sec:background}
The design of RNTuple\cite{EPJrntuple2020} comprises four layers: \textit{(i)} event iteration layer, that offers a convenient interface for looping over events; \textit{(ii)} logical layer, that maps complex C++ types onto columns; \textit{(iii)} primitives layer, which groups ranges of elements of a fundamental type into pages; and \textit{(iv)} storage layer, that is responsible for the I/O of pages, clusters, and required metadata.
This design makes it simple to support new data types or storage backends.

\begin{figure}[htb]
    \centering
    \resizebox{0.8\textwidth}{!}{\input{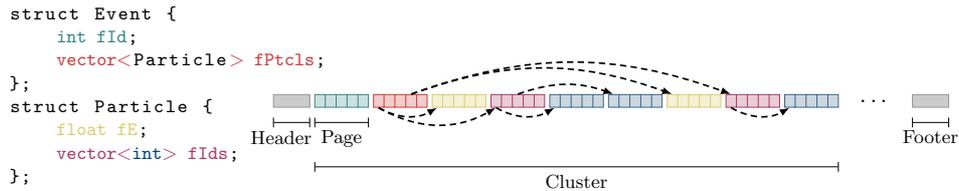}}
    \caption{RNTuple on-disk format. Pages store values for a specific data member (note the color coding).}\label{fig:rntuple_format}
\end{figure}

A page contains a certain range of values for a given column, whereas a cluster contains pages for a specific row range. Metadata includes a header that describes the data schema and a footer that contains the location of clusters and pages, among other information. Header/footer locations and their sizes are included in an anchor object.
Figure~\ref{fig:rntuple_format} shows a simplified example of the on-disk layout.
\vspace{-1em}

\vspace{-1.5em}
\section{Evaluation}\label{sec:evaluation}
In this section, we evaluate RNTuple w.r.t. TTree and other well-known I/O libraries outside HENP: HDF5 and Apache Arrow/Parquet.
Specifically, Section~\ref{sec:evaluation_qual} compares the support level of important HENP I/O layer features. A quantitative experimental evaluation is provided in Section~\ref{sec:evaluation_perf}.

To make a fair comparison, all our test programs were written in C++ and parameters such as row group size, page size, and compression algorithm and level were set to match in all cases, where permitted.
For Apache Parquet, we leveraged the Parquet-Arrow API permitting the convenient use of nested data structures and lists.
For HDF5, however, the columnar storage of heterogeneous data types or nested collections thereof is not a trivial problem. Parts of our test code bridge this gap and allows switching between alternate data layouts simply by changing a C++ template parameter.
Specifically, this layer provides the following layouts:

\begin{itemize}
    \item \textbf{Row-wise:} uses HDF5 compound types and variable-length types to represent nested structures and collections, respectively. This layout creates a single dataset whose type is the outer-most data structure and its dataspace dimension is $1\times N$.

    \item \textbf{Column-wise:} emulated columnar layout that uses one HDF5 dataset per column of a fundamental type, as described in~\cite{Sehrish:2017qxh}. Collections are translated to a HDF5 group and one additional index column.
\end{itemize}

In any case, HDF5 datasets are chunked. Given that chunks are individually accessed and (un)compressed, their size will be equal to the RNTuple or Parquet page size in all of our tests. The chunk cache size is set to the default size of a RNTuple cluster.

\vspace{-.4em}
\subsection{Qualitative evaluation}\label{sec:evaluation_qual}
In Table~\ref{tab:features}, we compare the level of support of different must-have features in a HENP I/O layer.
Compression is, in general, supported by all the analyzed storage formats; however, the native support for different algorithms greatly varies, e.g. HDF5 only supports zlib and szip.
Vertical and horizontal data combinations refers to extending the dataset with new entries or columns, respectively. To the best of our knowledge, Apache Arrow allows reading rows from many files, but it is unclear whether new columns can be made available.
Similarly, columnar access to multilevel nested structures and collections in HDF5 is unclear.
Schema evolution, i.e. handling changes in the data schema such as adding, removing or changing the type of a column, is only supported in TTree; preliminary support for this feature in RNTuple is foreseen for Q2 2022.
Native support for object stores is available in HDF5, RNTuple (DAOS), and Apache Arrow (S3). Finally, split encoding typically improves the compression ratio by reordering bytes in integer/floating point numbers, so that the $n-th$ byte of each value is contiguous in memory.

\begin{table}[t]
    \vspace{-1em}
    \centering
    {
\def\RotHdr#1{\rlap{\rotatebox{20}{#1}}\rule{1.5em}{0pt}}

\def\Yes{$\bullet$}
\def\Planned{{\color{black!25}$\bullet$}}
\def\Conditionally#1{{\color{lstkeyword}\textbf{(#1)}}}
\def\Partial{{\color{orange}\textbf{/}}}
\def\Unclear{{\color{orange}\textbf{?}}}

\scriptsize
\renewcommand{\arraystretch}{1.1}
\begin{tabular}{r@{~~}|cccc}
                                            & \RotHdr{HDF5} & \RotHdr{Parquet} & \RotHdr{TTree} & \RotHdr{RNTuple} \\
\hline
Transparent compression                     & \Conditionally{1} & \Yes      & \Yes  & \Yes \\
Columnar access                             & \Conditionally{2} & \Yes      & \Yes  & \Yes \\
Merging without uncompressing data          &                   &           & \Yes  & \Planned \\
Vertical/horizontal data combinations       &                   & \Partial  & \Yes  & \Planned \\
C++ and Python support                      & \Partial          & \Yes      & \Yes  & \Yes \\
Support for structs/nested collections      & \Unclear          & \Yes      & \Yes  & \Yes \\
Architecture-independent encoding           & \Yes              & \Yes      & \Yes  & \Yes \\
Schema evolution                            &                   &           & \Yes  & \Planned \\
Support for application-defined metadata    & \Yes              &           &       & \Planned \\
Fully checksummed                           & \Yes              & \Yes      &       & \Planned \\
Multi-threading friendly                    & \Planned          & \Yes      & \Yes  & \Yes \\
Native object-store support                 & \Yes              & \Yes      &       & \Yes \\
XRootD support                              &                   &           & \Yes  & \Yes \\
Automatic schema creation from C++ classes  &                   &           & \Yes  & \Yes \\
On-demand schema extension (backfilling)    &                   &           & \Yes  & \Planned \\
Split encoding / delta encoding             &                   & \Yes      &       & \Planned \\
Variable-length floats (min, max, bit size) &                   &           & \Yes  & \Planned \\
\end{tabular}\\[1em]
    
\tikz\node[text width=0.7\columnwidth,align=center,draw=black!7]{%
    $\bullet$~~Supported\hspace{1em}%
    \Planned~~Planned / Under development\hspace{1em}%
    \Partial~~Partial / Incomplete\hspace{1em}%
    \Unclear~~Unclear\\%
    \Conditionally{1}~~Only for chunked datasets\hspace{1em}%
    \Conditionally{2}~~Via emulated columnar};
}
\vspace{-1em}
    \caption{Comparison of features available in TTree, RNTuple, HDF5, and Apache Arrow/Parquet.}\label{tab:features}
    \vspace{-1em}
\end{table}

\vspace{-.4em}
\subsection{Experimental evaluation}\label{sec:evaluation_perf}
In this section, we provide experimental measurements of the analysis throughput, total amount of bytes read, and file size for TTree, RNTuple, HDF5 (both row-wise and column-wise) and Apache Parquet in a variety of situations.
The hardware and software environment, datasets used, and test cases are described in the following.

\vspace{-0.5em}
\paragraph{Hardware and software environment.}
Our benchmarks ran on a single node based on $1\times$AMD EPYC 7702P 64-Core processor running at 2\,GHz, and 128\,GB DDR4 RAM. SMT was enabled, although disabling it yielded similar results for the workload in our tests.
This machine is also equipped with a Samsung PM173X NVMe SSD, and a TOSHIBA MG07ACA1 SATA hard disk drive. A ext4 filesystem resides on each drive using a 4\,KB block size and default mount options.
CephFS was used as the network filesystem for tests that operate over a network share.

The software environment is based on CentOS Linux 8.3 (kernel 5.15.1), Apache Arrow 5.0.0, HDF5 1.10.5, and ROOT git revision \texttt{5001281762} built with g++ 8.4.1.

\vspace{-0.5em}
\paragraph{Test cases.}
The experiments ran in the evaluation have used the following datasets as input:
\begin{itemize}
    \item \textbf{LHCb Run 1 Open Data B2HHH (B meson decays to three hadrons).} No nested collections, containing 8.5\,M events, 26 branches\footnote{In TTree, the term ``branch'' refers to a column; usually, both terms can be used interchangeably.}. The compressed file size is 1.1\,GB.
    \item \textbf{CMS Open Data Higgs $\to$ 4 leptons MC.} NanoAOD-like\cite{EPJnanoaod2019} format, 300\,k events, 84 branches. This dataset was concatenated 16 times so as to make a larger file of $\sim2.1$\,GB.
\end{itemize}

We carried out two different experiments\cite{iotoolsACAT21}: \textit{(i)} running a simple analysis program over the LHCb dataset that generates the B mass spectrum histogram and measures the end-to-end, i.e. from storage to histogram, analysis throughput in uncompressed MB/s; and \textit{(ii)} measurement of the read rate in uncompressed MB/s for both the LHCb and the CMS datasets, retrieving all entries in 10 selected branches.

The original files were in TTree format. In a first stage, a third program is used to generate the equivalent files for each of the other formats (HDF5 row-wise, HDF5 column-wise, and Apache Parquet).
All files taken as input by the benchmark programs use zstd with compression level 5, except for HDF5 that uses zlib with compression level 3 given that HDF5 lacks native support for zstd.
RNTuple and Apache Parquet multi-threaded I/O and (un)compression was enabled in all cases.
RNTuple benchmarks use a cluster prefetch value of $5$.

Each experiment and storage format combination is run four times, differing on the location of the input file: CephFS, SSD, HDD. To ensure that data is forcibly read from the underlying storage, the Linux page cache is cleared prior to running each test. A fourth iteration uses the same input file as in the HDD case but does not clear the page cache before the test is executed.
Additionally, we measured the total file size after conversion and the raw bytes read by each experiment--format pair, as reported by the vmtouch\cite{doug2012vmtouch} tool.
In all the plots shown below, the data point and error bars (where applicable) refer to the average and minimum/maximum values measured over 10 executions, respectively.

\begin{figure}[htb]
    \centering
    \vspace{-0.5em}
    \input{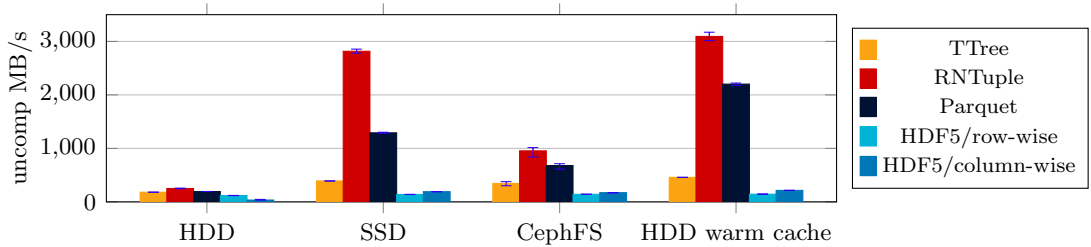}\vspace{-1em}
    \caption{LHCb B2HHH analysis throughput (18/26 branches; compressed).}\label{fig:b2hhh_perf}
    \vspace{-.5em}
\end{figure}

In a first stage, we measured the analysis throughput for all combinations of format and storage device for the LHCb dataset. The analysis program requires reading 18 out of 26 columns.
As can be seen in Figure~\ref{fig:b2hhh_perf}, RNTuple speedup over Apache Parquet is between $\times$1.4 and $\times$2.2 depending on the scenario. Also note that the performance of HDF5 is severely affected by the lack of multi-threaded I/O and decompression.

\begin{figure}[htb]
    \centering
    \input{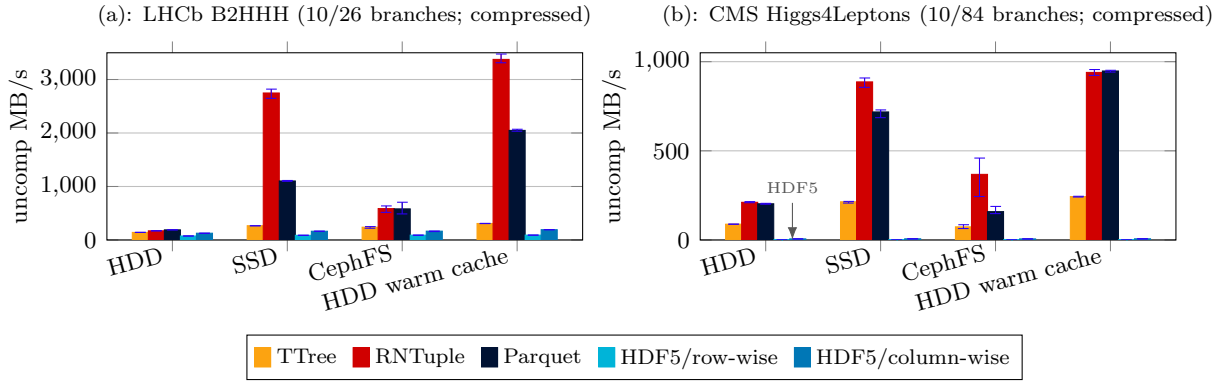}\vspace{-2em}
    \caption{Read rate (uncompressed MB/s; 10 selected branches). Note the low ($<10$\,MB/s) transfer rate for HDF5 in the CMS case.}\label{fig:partialread}
\end{figure}

Whereas the previous plots give an idea of the performance of each candidate, these results do not represent the behavior in case of accessing collections.
In a second experiment, we measured the read rate in uncompressed MB/s for 10 columns of both the LHCb and CMS dataset. This test allows us to compare the performance differences for collections.
As can be seen in Figure~\ref{fig:partialread}, for the LHCb dataset, RNTuple's worst result is comparable to Apache Parquet. In all the other tests, RNTuple outperforms other alternatives.
For the CMS dataset, RNTuple worst case achieves at least the same result as Parquet. Differences between both plots is explained by the use of different column types (\texttt{double} w.r.t. \texttt{float}) and the presence of collections.

\begin{figure}[htb]
    \centering
    \scalebox{0.87}{\tikz{\begin{axis}[xbar=0,ytick=data,enlargelimits=0.1,enlarge y limits=0.2,
    bar width=4.75pt,y=14pt,
    width=\columnwidth,height=6in,
    xlabel={GB ($10^9$ bytes)},
    yticklabels from table={data/size_readb_comp.csv}{[index]0},
    y dir=reverse,
    cycle list name=twobars,
    legend style={nodes={scale=0.8,transform shape}},
    legend image code/.code={\node[fill,draw,inner sep=3pt] {};},
    label style={font=\footnotesize},tick label style={font=\footnotesize},
    xtick={0.25,0.5,...,3.25},xmin=0.3,xmax=3.3,
    x tick label style={
        /pgf/number format/.cd,fixed,fixed zerofill,precision=2
    }
]
    \addplot table [domain=0.3:3.3,x=readb,y expr=\coordindex] {data/size_readb_comp.csv};
    \addlegendentry{Read bytes}

    \addplot table [domain=0.3:3.3,x=size,y expr=\coordindex] {data/size_readb_comp.csv};
    \addlegendentry{File size}
\end{axis}}}\vspace{-1.25em}
    \caption{File size and raw bytes read for the CMS dataset (10/84 branches).}\label{fig:size_readb_comp}
    \vspace{-1.25em}
\end{figure}
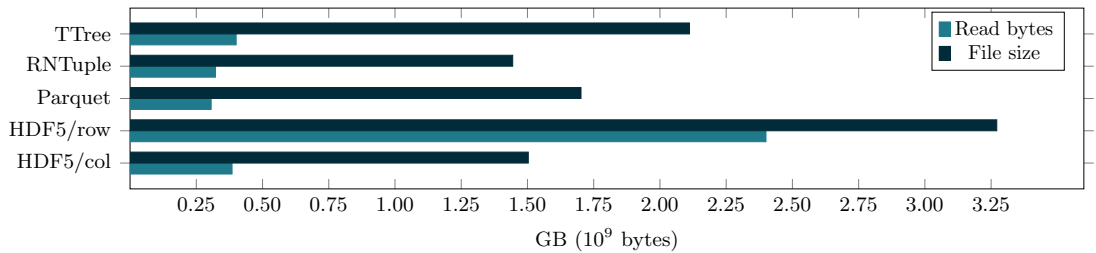

Finally, we measured the compressed file size and total bytes transferred during the last test for the CMS dataset (see Figure~\ref{fig:size_readb_comp}). As can be seen, RNTuple provides the smallest file while the amount of bytes read is about the same as Parquet.
It is worth noting that HDF5 row-wise does not read the whole file if the compound type definition provided at run-time misses some members w.r.t. the committed (on-file) type; however, the throughput in this case is extremely low ($<1$\,MB/s; see Figure~\hyperref[fig:partialread]{\ref*{fig:partialread}.b}).

\vspace{-0.5em}
\section{Conclusion and future work}\label{sec:conclusion}
RNTuple is able to deliver the highest throughput among the analyzed alternatives in all of our tests thanks to its performance-tuned implementation and parallel I/O scheduling and decompression. The latest developments are available in the \texttt{ROOT::Experimental} namespace. The feature roadmap aims at complete support for the HENP event data storage use case.

RNTuple is expected to become production grade in 2024. A number of important features are scheduled for 2022: schema evolution, on-demand addition of new columns to the model, complete support for vertical/horizontal data combinations, and merging without uncompressing pages, only to name a few.
As a future work, we also plan to compare the performance achieved by the RNTuple and HDF5 DAOS connectors.

\vspace{-0.7em}
\section*{Acknowledgments}
This work benefited from support by the CERN Strategic R\&D programme on Technologies for Future Experiments (CERN-OPEN-2018-06).

\vspace{-0.7em}
\printbibliography
\end{document}